\documentclass[aps,prl,twocolumn,superscriptaddress,amsmath,amssymb,floatfix,longbibliography]{revtex4-2}

\usepackage{graphicx}
\usepackage{bm}
\usepackage{mathrsfs}
\usepackage[colorlinks=true,linkcolor=blue,citecolor=blue,urlcolor=blue]{hyperref}

\newcommand{\B}{\mathfrak{B}}
\newcommand{\Cl}{\mathscr{C}}
\newcommand{\GPJ}{G_{\rm PJ}}
\newcommand{\GH}{G_{\rm H}}
\newcommand{\GPV}{G_{\rm PV}}
\newcommand{\rmd}{\mathrm{d}}
\newcommand{\Tr}{\operatorname{Tr}}
\newcommand{\Img}{\operatorname{Im}}

\begin{document}

\title{Bargmann-invariant witness of the graviton commutator}

\author{Hiroki Matsui}
\email{hiroki.matsui@yukawa.kyoto-u.ac.jp}
\affiliation{Osaka Central Advanced Mathematical Institute (OCAMI), Osaka Metropolitan University, Osaka 558-8585, Japan}
\affiliation{TRIP Headquarters, RIKEN, Wako, Saitama 351-0198, Japan}

\date{\today}

\begin{abstract}
We identify the phase of a Bargmann invariant of two matter-wave interferometers as a quantum gravity signature that cannot be reproduced by any branchwise-definite classical gravitational field. For matter coupled linearly to linearized gravity, the interferometric coherence factorizes exactly into a conservative Newtonian phase, the characteristic functional of the graviton state, and a commutator phase. For Wigner-positive graviton states, the first two are classically reproducible; therefore, decoherence alone cannot certify that gravity is quantized. The Bargmann phase isolates the commutator, which is independent of the graviton state and of all Newtonian interactions and is bounded by which-path decoherence.
For gravity, this radiation-reaction effect is sixty orders of magnitude below experimental reach, but trapped-ion analogs can test it at order unity.
\end{abstract}

\maketitle

\textit{Introduction.}---Proposals for experiments designed to demonstrate the quantum nature of gravity are based on the idea of gravitationally mediated entanglement between two masses in spatial superposition~\cite{Bose:2017nin,Marletto:2017kzi}. Opinions are divided as to what these experiments would actually establish. In the Newtonian regime, the entangling phase can be reproduced by assigning a definite classical gravitational field to each quantum branch of the matter sources~\cite{Christodoulou:2018cmk,Martin-Martinez:2022uio}. The roles of propagating gravitational degrees of freedom, retardation, and quantum matter sources in gravitational quantum effects have been investigated~\cite{Belenchia:2018szb,Danielson:2021egj,Christodoulou:2022mkf,Mitrakos:2026xlr,Fujita:2026upv}, and the effects of decoherence, vacuum fluctuations, and causality in field-mediated interactions have been analyzed in Refs.~\cite{Sugiyama:2022wcd,Hidaka:2022tzk,Hidaka:2022gsv,Sugiyama:2023tqu}. Related no-go theorems, hybrid classical--quantum models, and entanglement-free tests have further clarified the question~\cite{Galley:2020qsf,Lami:2023gmz,Hanif:2023fto,Aziz:2025ypo,Oppenheim:2018igd,Oppenheim:2022xjr,Fabiano:2026agi}. Complementary studies have investigated the decoherence of matter caused by gravitons~\cite{Blencowe:2012mp,Anastopoulos:2013zya,Oniga:2015lro,Kanno:2020usf,Matsui:2026skx}, and it has been argued that amplified graviton noise in a squeezed or relic background would signal quantization~\cite{Kanno:2021gpt,Parikh:2020kfh}. However, decoherence in a single system can generally be mimicked by classical noise~\cite{Helm:2009uhf,Gu:2019vcm}, and similar caveats apply to the detection of gravitons~\cite{Carney:2023nzz,Tobar:2023ksi}.

In this Letter, we ask which features of matter-wave interference cannot be explained by classical gravity. We consider models in which each branch experiences a definite classical gravitational field, allowing for classical noise and branch-dependent phases. Building on the exact solution of branch-dependent graviton emission~\cite{Matsui:2026skx}, we show that the matter-wave coherence between any two branches factorizes into three contributions: a conservative gravitational phase, a factor describing field fluctuations, and a phase arising from the field commutator. For graviton states with nonnegative Wigner functionals, the first two contributions can be reproduced by classical branch-dependent phases and noise. We then show that a four-cycle Bargmann invariant constructed from two interferometers removes these contributions from its phase and isolates the commutator, independently of the graviton state and of all Newtonian interactions. An uncertainty relation for displacement operators further bounds this signal in terms of which-path decoherence. Finally, we estimate its magnitude for gravity and discuss a bosonic-mode analog.

\textit{Branch evolution.}---Matter is prepared in a superposition of orthonormal branch states $|a\rangle$, each sourcing a classical conserved stress tensor $T_a^{\mu\nu}$. Histories are closed: branches coincide before splitting and after recombination, and apparatus recoil is included, so the difference sources $T_a-T_b$ carry no net energy or momentum~\cite{Matsui:2026skx}. With $g_{\mu\nu}=\eta_{\mu\nu}+\kappa h_{\mu\nu}$, $\kappa^2=32\pi G$, and $H_{\rm int}=-\frac{\kappa}{2}\int\rmd^3x\,h_{\mu\nu}T^{\mu\nu}$, write $J^{\mu\nu}\equiv(\kappa/2)T^{\mu\nu}$, let $h[J]\equiv\int\rmd^4x\,h_{\mu\nu}J^{\mu\nu}$ denote the source-smeared interaction-picture field, and define, for any kernel $K$,
\begin{equation}
(T|K|T')\equiv\frac{\kappa^2}{4}\int\rmd^4x\,\rmd^4x'\,T^{\mu\nu}(x)K_{\mu\nu,\rho\sigma}(x,x')T'^{\rho\sigma}(x')
\label{eq:bracket}
\end{equation}
(we set $\hbar=c=1$ except in the paragraph on gravity). The field commutator is a $c$-number, $[h(x),h(x')]=-i\GPJ(x,x')$, with $\GPJ=i\langle[h,h]\rangle$ the Pauli--Jordan kernel; it decomposes into the retarded and advanced Green functions, $G_R=\theta(t-t')\GPJ$ and $G_A(x,x')=G_R(x',x)$, so that, by the Peierls bracket~\cite{Peierls:1952cb},
\begin{equation}
[h[J],h[J']]=-i(T|\GPJ|T'),\qquad \GPJ=G_R-G_A,
\label{eq:comm}
\end{equation}
where $(T|G_R|T')=(T'|G_A|T)$ is the retarded field radiated by $T'$ evaluated on $T$. Because the coupling is linear, the Magnus expansion terminates~\cite{Matsui:2026skx} and the evolution in branch $a$ is
\begin{equation}
U_a=e^{iS_a}e^{ih[J_a]},\qquad S_a=\tfrac12(T_a|G_R|T_a),
\label{eq:Ua}
\end{equation}
where only the time-symmetric part $\GPV\equiv\tfrac12(G_R+G_A)$ of the response contributes to $S_a$, and self-energies are absorbed into the matter parameters. Tracing out the gravitons and using the Baker--Campbell--Hausdorff formula gives the matter coherences $\rho_{ab}=\rho^{(0)}_{ab}M_{ab}$, with
\begin{align}
M_{ab}&=e^{i(S_a-S_b)}\,e^{\frac{i}{2}(T_a|\GPJ|T_b)}\,\chi[T_a-T_b],\nonumber\\
\chi[T]&\equiv\Tr\big(\rho_h e^{ih[J]}\big).
\label{eq:master}
\end{align}
Equation~\eqref{eq:master} is exact for any graviton state $\rho_h$ and any number of branches; for two branches it reduces to the result of Ref.~\cite{Matsui:2026skx}, where the middle factor is absorbed into the relative phase. The two phases combine into the Feynman--Vernon response phase $(\Delta_{ab}|G_R|\Sigma_{ab})$, with $\Delta_{ab}\equiv T_a-T_b$ and $\Sigma_{ab}\equiv\tfrac12(T_a+T_b)$ (Supplemental Material, Sec.~\ref{sec:master}). The three factors of $M_{ab}$ play distinct roles. The phase $S_a$ depends only on $\GPV$, whose Fourier transform is the principal value of $1/k^2$ with no $\delta(k^2)$ part; it is off shell and contains the Newtonian potential and all conservative post-Newtonian (PN) interactions, both within and between subsystems. The factor $\chi$ is the characteristic functional of the graviton state, evaluated at the difference source $\Delta_{ab}$; for the vacuum, $|\chi[\Delta_{ab}]|=e^{-\Gamma_{\rm vac}}$ with $\Gamma_{\rm vac}=\tfrac12(\Delta_{ab}|\GH|\Delta_{ab})$, where $\GH=\tfrac12\langle\{h,h\}\rangle$ is the Hadamard (noise) kernel; $\Gamma_{\rm vac}$ is half the mean number of gravitons radiated by $\Delta_{ab}$~\cite{Matsui:2026skx}. Finally, $\GPJ$ is state independent. Its Fourier transform is supported on $k^2=0$, where for conserved sources the de~Donder polarization sum collapses to the two transverse-traceless helicities~\cite{Weinberg:1972kfs}; hence $(T|\GPJ|T')$ is gauge invariant and determined by on-shell gravitons alone,
\begin{equation}
(T|\GPJ|T')=2\Img\langle\alpha_{T'},\alpha_{T}\rangle,
\label{eq:Ealpha}
\end{equation}
where $\alpha_T$ is the coherent radiation amplitude sourced by $T$ and $\langle\alpha,\beta\rangle=\sum_s\int\rmd^3k\,(2\pi)^{-3}\alpha_s^*\beta_s$, as in Ref.~\cite{Matsui:2026skx} (Sec.~\ref{sec:master}).

\textit{Branchwise-definite models.}---We define $\Cl$ as the class of models in which each matter branch $a$ acquires a phase $\theta_a=s_a+h_{\rm cl}[T_a]$. Here $s_a$ is an arbitrary deterministic phase that may depend on the sources of all branches and on the quantum state of the matter. The second term is the phase imparted by a classical random field $h_{\rm cl}$ coupled linearly to $T_a$. A single realization of this field acts on all branches, and its probability distribution $P$ does not depend on the branch. Averaging over this field gives
\begin{equation}
M^{\Cl}_{ab}=e^{i(s_a-s_b)}\,\Phi_P[T_a-T_b],\qquad \Phi_P[T]\equiv\mathbb{E}_P\big[e^{ih_{\rm cl}[T]}\big].
\label{eq:classical}
\end{equation}
This class covers the phase and noise effects of semiclassical (mean-field) gravity~\cite{Page:1981aj}, Newtonian action at a distance, quantum-controlled classical fields~\cite{Christodoulou:2018cmk,Martin-Martinez:2022uio}, classical stochastic gravitational-wave backgrounds, and Di\'osi--Penrose-type gravitational noise~\cite{Diosi:1989hlx,Penrose:1996cv,Tilloy:2015zya}. In particular, the Newtonian entangling phase of Refs.~\cite{Bose:2017nin,Marletto:2017kzi} can be reproduced through $s_a$. The class excludes dynamics in which the field responds to a coherence $\rho_{ab}$ rather than to the individual branches; we return to such hybrid models below.

\textit{Theorem 1} (classical simulability). If the Wigner functional $W$ of the graviton state is nonnegative and $(T_a-T_c|\GPJ|T_b-T_c)=0$ for all branches $a,b,c$, then all branch coherences $\{M_{ab}\}$ can be reproduced by a model in $\Cl$. The commutator condition always holds for a single two-path interferometer, and $W\geq0$ for the vacuum, thermal, coherent, and squeezed states, the inflationary relic background~\cite{Grishchuk:1990bj}, and their mixtures.

\emph{Proof sketch.} A nonnegative $W$ is a probability distribution for the initial data of a free classical field $h_{\rm cl}$, and averaging $e^{ih_{\rm cl}[T]}$ over it gives $\chi[T]=\Phi_W[T]$. The commutator condition implies $\tfrac12(T_a|\GPJ|T_b)=f_a-f_b$ for some real $f_a$ (Sec.~\ref{sec:theorem1}), so Eq.~\eqref{eq:master} takes the form of Eq.~\eqref{eq:classical} with $P=W$ and $s_a=S_a+f_a$.~$\square$

Consequently, the decoherence of a single interferometer\nobreak\hspace{0pt}---\allowbreak$\Gamma_{\rm vac}$, its squeezing-dependent generalization, and the inflationary enhancement $\Gamma_{\rm inf}$ of Ref.~\cite{Matsui:2026skx}---is reproduced exactly by classical stochastic linearized gravity with noise drawn from $W$, in line with the corresponding statement for inflationary perturbations~\cite{Martin:2015qta} and with positive-Wigner simulability~\cite{Mari:2012ypq}. For Wigner-positive states, the only obstruction to $\Cl$ is the commutator between \emph{distinct} difference sources.

\begin{figure}[t]
\centering
\includegraphics[width=\columnwidth]{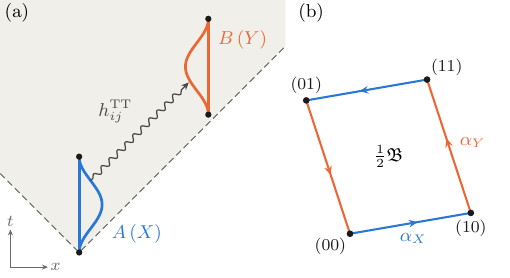}
\caption{(a) Two closed interferometers $A$ and $B$ with difference sources $X$ and $Y$; $B$ lies in the causal future of $A$, so on-shell gravitons (wavy line) connect them. (b) The four joint branches in graviton phase space. The witness~\eqref{eq:Bdef} traverses the cycle $(00)\to(10)\to(11)\to(01)$ and equals twice the enclosed symplectic area, $\B=(X|\GPJ|Y)=2\Img\langle\alpha_Y,\alpha_X\rangle$; in every branchwise-definite model $\B=0$.}
\label{fig:setup}
\end{figure}

\textit{Bargmann witness.}---Consider two interferometers $A$ and $B$ with difference sources $X\equiv T^A_1-T^A_0$ and $Y\equiv T^B_1-T^B_0$, so that the joint branches carry $T_{ab}=T_{00}+aX+bY$ (Fig.~\ref{fig:setup}). For an initially uncorrelated state, e.g., $|{+}\rangle_A|{+}\rangle_B$, we consider the four-cycle Bargmann invariant~\cite{Bargmann:1964zj,PhysRevLett.70.880}, whose phase has also been used to detect the imaginarity of sets of states~\cite{Fernandes:2024sik}, and define its phase
\begin{equation}
\B\equiv\arg\big(\rho_{00,10}\,\rho_{10,11}\,\rho_{11,01}\,\rho_{01,00}\big),
\label{eq:Bdef}
\end{equation}
which is invariant under local rephasings $|a\rangle\to e^{i\lambda_a}|a\rangle$ and can be measured with conditional Ramsey fringes or cycle tests~\cite{Oszmaniec:2021rzh}. Substituting Eq.~\eqref{eq:master} into Eq.~\eqref{eq:Bdef}, we find that the phases $S_a-S_b$ cancel around the cycle. The factors $\chi$ combine into $\chi[-X]\chi[-Y]\chi[X]\chi[Y]=|\chi[X]\chi[Y]|^2\geq0$, since $\chi[-T]=\chi[T]^*$, and therefore contribute no phase. The remaining commutator terms give a total phase $(X|\GPJ|Y)$:
\begin{equation}
\B_{\rm Q}=(X|\GPJ|Y)=2\Img\langle\alpha_Y,\alpha_X\rangle .
\label{eq:BQ}
\end{equation}
In every model of $\Cl$, Eq.~\eqref{eq:classical} gives instead
\begin{equation}
\B_{\Cl}=\arg\big(|\Phi_P[X]\,\Phi_P[Y]|^2\big)=0,
\label{eq:BC}
\end{equation}
for every distribution $P$, Gaussian or not, and every choice of $s_a$. Equations~\eqref{eq:BQ} and \eqref{eq:BC} are our central result. A nonzero $\B$ excludes the entire class $\Cl$; it is independent of the graviton state, blind to Newtonian and all other conservative interactions, gauge invariant, and fixed by on-shell gravitons. The witness $\B$ vanishes when the two interferometers are spacelike separated, because the field commutator vanishes outside the light cone. It also vanishes for time-symmetric protocols, in which both difference sources are invariant under time reversal about a common instant, since $\GPJ$ changes sign under time reversal. These include the simultaneous, time-symmetric splitting of two masses. The derivation assumes that the sources are additive, $T_{ab}=T_{00}+aX+bY$. Mutual gravitational forces modify the trajectories and violate this assumption at $O(G)$. The resulting corrections can make $\B_{\Cl}$ nonzero only at higher order in the gravitational coupling and only for non-Gaussian noise (Sec.~\ref{sec:witness}).

The witness has a transparent operational meaning. Let $\phi_{A|b}\equiv\arg\rho_{1b,0b}$ be the fringe phase of $A$ when $B$ is in path $b$, and $\delta\phi_{A|B}\equiv\phi_{A|1}-\phi_{A|0}$ the shift that an experimenter operating $A$ observes when $B$ is switched from path $0$ to path $1$; define $\delta\phi_{B|A}$ likewise. Then $\B=\delta\phi_{A|B}-\delta\phi_{B|A}$. In $\Cl$, $s_a$ is a property of the joint branch, so the two conditional shifts coincide: phase kickback is reciprocal. Equation~\eqref{eq:master} gives instead
\begin{equation}
\delta\phi_{A|B}=(X|G_R|Y),\qquad \delta\phi_{B|A}=(Y|G_R|X),
\label{eq:shifts}
\end{equation}
for any $\rho_h$: each fringe responds to the retarded field of the other difference source, as in the two-qubit model of Ref.~\cite{Hidaka:2022gsv}. If $A$ lies entirely in the causal past of $B$, then $\delta\phi_{A|B}=0\neq\delta\phi_{B|A}$---the influence is one way. A branchwise-definite model cannot reproduce this: matching the observed $\delta\phi_{B|A}$ forces the same shift on $A$, so the beam-splitter setting of $B$ would change the fringe visibility of $A$ even when $A$ is read out at spacelike separation---the signaling tension identified in Refs.~\cite{Belenchia:2018szb,Danielson:2021egj}. The commutator term in Eq.~\eqref{eq:master} is what keeps the quantum description causal, and $\B$ measures it directly.

\textit{Quantumness budget.}---The commutator signal is not free. We apply the uncertainty relation for unitary operators~\cite{Massar:2007akt,Bagchi:2016igu}, $\Delta U\Delta V\geq\frac12|\langle[U,V]\rangle|$ with $\Delta U^2=1-|\langle U\rangle|^2$, to $U=e^{ih[J_X]}$ and $V=e^{\pm ih[J_Y]}$. The commutation relations of these displacement operators give $|\langle[U,V]\rangle|=2|\sin(\B/2)|\,|\chi[X\pm Y]|$, so that, for every graviton state,
\begin{equation}
|\sin(\B/2)|\,V_{X\pm Y}\leq\sqrt{(1-V_X^2)(1-V_Y^2)},
\label{eq:budget}
\end{equation}
where $V_Z\equiv|\chi[Z]|$ are visibilities read off the same data set: $V_X$ and $V_Y$ from $|\rho_{00,10}|$ and $|\rho_{10,11}|$, and $V_{X\pm Y}$ from $|\rho_{00,11}|$ and $|\rho_{10,01}|$, each normalized to its initial value. With $\Gamma_Z\equiv-\ln V_Z$, Eq.~\eqref{eq:budget} becomes, in the weak-coupling regime,
\begin{align}
\B^2&\leq16\,\Gamma_X\Gamma_Y,\nonumber\\
|\B|&\leq2\|\alpha_X\|\|\alpha_Y\|=4\sqrt{\Gamma^{\rm vac}_X\Gamma^{\rm vac}_Y},
\label{eq:budget2}
\end{align}
where the second inequality follows from Eq.~\eqref{eq:BQ} by the Cauchy--Schwarz inequality and is saturated for $\alpha_Y\propto i\alpha_X$. The first inequality is the Robertson-type bound relating causal response to decoherence, derived in Ref.~\cite{Sugiyama:2022wcd} for one-way response, where $\B=-\delta\phi_{B|A}$, and in Ref.~\cite{Hidaka:2022gsv}, with the noise cross-correlation included, for the difference of the two retarded responses; Eq.~\eqref{eq:budget} extends it to arbitrary graviton states and finite coupling. One-way influence through a field therefore requires that the field record who influenced whom: each interferometer must leak which-path information into on-shell gravitons at a comparable level. Three consequences follow. First, squeezed or relic graviton backgrounds change $\Gamma_{X,Y}$ but leave $\B$ unchanged; the relic background, which enhances $\Gamma_{X,Y}$~\cite{Matsui:2026skx}, only reduces the visibility $V_XV_Y$ on which the phase estimate relies. Second, Eq.~\eqref{eq:budget} relies on relations among branch coherences imposed by the displacement operators, rather than on positivity of $\rho$ alone. Generic positive four-branch coherence matrices can violate this bound (Sec.~\ref{sec:budget}), so it is an independent consistency test of the quantum-field description. Third, measuring the witness $\B$ is much more demanding than detecting decoherence. In the weak-coupling regime, where the visibilities are close to unity, resolving the small phase $\B$ requires $N\sim\B^{-2}$ repetitions (Sec.~\ref{sec:witness}). Since $\B^2\leq16\Gamma_X\Gamma_Y$ by Eq.~\eqref{eq:budget2}, this implies $N\gtrsim(16\Gamma_X\Gamma_Y)^{-1}$, whereas ideally $N\sim\Gamma^{-1}$ repetitions suffice to detect the decoherence.

\begin{figure}[t]
\centering
\includegraphics[width=\columnwidth]{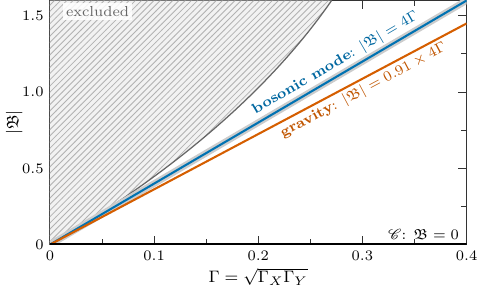}
\caption{Quantumness budget in the plane of which-path decoherence $\Gamma\equiv\sqrt{\Gamma_X\Gamma_Y}$ and witness $|\B|$ (radians). Hatched: excluded by Eq.~\eqref{eq:budget}, drawn for $V_X=V_Y=e^{-\Gamma}$ and $V_{X\pm Y}=V_XV_Y$; the gray line $|\B|=4\Gamma$ is its weak-coupling form, Eq.~\eqref{eq:budget2}. Every branchwise-definite model $\Cl$ lies on the axis $\B=0$, whatever its decoherence. Quantum predictions saturate the budget at leading order: a ground-state bosonic mode kicked in orthogonal quadratures (trapped-ion analog) gives $|\B|=2|\alpha_X||\alpha_Y|=4\Gamma$ exactly (blue), and the gravitational model~\eqref{eq:gauss} gives $0.91\times4\Gamma$ for parallel axes at $\delta=0.62$ (vermilion; $\Gamma\lesssim10^{-60}$ for proposed experiments, Sec.~\ref{sec:grav}). Thermal or relic backgrounds and hybrid classical--quantum dynamics increase $\Gamma$ but leave $\B$ unchanged, moving a prediction into the interior; decoherence alone therefore cannot certify the commutator.}
\label{fig:budget}
\end{figure}

\textit{Gravity.}---For nonrelativistic sources whose separation $L$ obeys $L\ll c\tau$, with $\tau$ the protocol time scale, the difference sources have vanishing mass monopole and dipole once recoil is included, and the quadrupole limit of Eq.~\eqref{eq:BQ} gives (Sec.~\ref{sec:grav})
\begin{equation}
\B=\frac{2G}{5\hbar c^5}\int\rmd t\,\Delta Q^B_{ij}(t)\,\frac{\rmd^5\Delta Q^A_{ij}}{\rmd t^5},
\label{eq:quad}
\end{equation}
where $\Delta Q^{A,B}_{ij}$ are the trace-free quadrupole differences. This is the cross term of the Burke--Thorne radiation-reaction potential~\cite{Burke:1970wx,Thorne:1969rba,Blanchet:2013haa}, which enters at the first time-odd (2.5PN) order; it is antisymmetric under $A\leftrightarrow B$ and independent of $L$. Time-symmetric retardation corrections at 1PN and 2PN enter only $S_a$. For the Gaussian trajectory model of Ref.~\cite{Matsui:2026skx}---masses $m_{A,B}$, branch separation $d_{A,B}$ along unit axes $\bm e_{A,B}$, time scale $\tau$, and $B$ delayed by $\Delta t=\delta\tau$---Eq.~\eqref{eq:quad} evaluates to
\begin{equation}
\B=-\frac{2Gm_Am_Bd_A^2d_B^2}{5\hbar c^5\tau^4}\Big[(\bm e_A\!\cdot\bm e_B)^2-\tfrac13\Big]\,I(\delta),
\label{eq:gauss}
\end{equation}
with $I(\delta)=\sqrt{\pi/2}\,(\delta^5-10\delta^3+15\delta)\,e^{-\delta^2/2}$. Since $\Gamma_{\rm vac}=\frac{8}{15}Gm^2d^4/\hbar c^5\tau^4$ for each interferometer~\cite{Matsui:2026skx}, the budget ratio $|\B|/4\sqrt{\Gamma^{\rm vac}_X\Gamma^{\rm vac}_Y}=\frac{3}{16}|(\bm e_A\!\cdot\bm e_B)^2-\frac13|\,|I(\delta)|$ reaches $0.906$ for parallel axes at $\delta\simeq0.62$ (Fig.~\ref{fig:budget}): gravitational radiation nearly saturates the budget. In this model identical interferometers give $|\B|\leq3.62\,\Gamma_{\rm vac}$, close to the general bound $4\Gamma_{\rm vac}$. With the optimistic parameters of Ref.~\cite{Matsui:2026skx}, $m=10^{-14}\,$kg, $d=250\,\mu$m, $\tau=0.5\,$s, one finds $|\B|_{\max}\simeq3.2\times10^{-60}$, requiring of order $10^{119}$ repetitions; other platforms considered there give $10^{-88}$--$10^{-69}$ (Sec.~\ref{sec:grav}).

\textit{Analog test.}---The same displacement-operator algebra underlies geometric phase gates in trapped ions~\cite{Soerensen:1998mp,Leibfried:2003rkv,Hou:2024bcs} and provides a tabletop analog of the witness. Let two qubits $A$ and $B$ share a motional mode, with the qubit states playing the role of the branches and the mode that of the field. A spin-dependent kick conditioned on $A$ displaces the mode by $\alpha_X$; a later kick conditioned on $B$ displaces it by $\alpha_Y$ in the orthogonal quadrature. If the displacements are left open, so that the mode keeps a record of both kicks, Eq.~\eqref{eq:BQ} predicts $\B=2\Img(\alpha_Y^*\alpha_X)$ for any motional state, while the qubit visibilities are set by the characteristic function of that state and decrease with its thermal occupation---the analog of which-path decoherence. A classical thermal oscillator reproduces this decay but, belonging to $\Cl$, gives $\B=0$ at every temperature. The gates themselves close the displacements into loops; this erases the record in the mode---the analog of on-shell gravitons---and leaves only a reciprocal entangling phase of the $S_a$ type, the analog of Newtonian entanglement. For $|\alpha|\simeq0.5$, one expects $|\B|\sim0.5$ and near-saturation of the budget~\eqref{eq:budget} for a ground-state mode, so both the witness and the budget are testable with current technology.

\textit{Discussion.}---Previous work showed that Newtonian entanglement can be reproduced by quantum-controlled classical fields~\cite{Christodoulou:2018cmk,Martin-Martinez:2022uio} and emphasized the role of retardation in probing the quantum nature of the gravitational field~\cite{Danielson:2021egj,Christodoulou:2022mkf,Mitrakos:2026xlr}. Our results identify a concrete observable: the difference $\B$ between the phase shifts that the two interferometers induce in each other. In the quantum theory, this difference is determined by the field commutator and depends only on on-shell gravitons. A nonzero $\B$ cannot be reproduced by any model in $\Cl$. Equation~\eqref{eq:budget} relates this signal to which-path decoherence, while Eq.~\eqref{eq:budget2} bounds its magnitude in terms of the vacuum decoherence exponents associated with graviton emission. For the proposed experiments considered in Ref.~\cite{Matsui:2026skx}, this limits $|\B|$ to at most order $10^{-60}$ radians, far below foreseeable experimental sensitivity. Squeezed or relic backgrounds can change the visibility but cannot increase $\B$, because the field commutator is independent of the graviton state.

These conclusions have two limitations. First, a nonzero $\B$ excludes every branchwise-definite model but not the hybrid classical--quantum dynamics of Refs.~\cite{Aleksandrov+1981+902+908,Boucher:1988ua,Diosi:2014ebu,Oppenheim:2018igd}. There the classical field within each coherence $\rho_{ab}$ is driven by the mean source $\tfrac12(T_a+T_b)$, as dictated by the Aleksandrov--Boucher--Traschen bracket, and this reproduces the full phase of Eq.~\eqref{eq:master}, including $\B$ (Sec.~\ref{sec:budget}). Such dynamics remain positive only when supplemented by decoherence and diffusion~\cite{Oppenheim:2022xjr,Layton:2023oud}, so they differ from quantum gravity in the amount of noise they require rather than in $\B$; confronting the budget~\eqref{eq:budget} with their decoherence--diffusion trade-off is a natural next step. Second, if the graviton state had a negative Wigner functional, $\chi$ could not be the characteristic functional of any classical field (Bochner's theorem), and measurements on a single interferometer would already exclude $\Cl$. No known mechanism, however, prepares such states of the radiative field. Extending the analysis to Killing horizons~\cite{Danielson:2022sga} and to soft-graviton backgrounds~\cite{Matsui:2026abk} is left for future work.

\textit{Data availability.}---The numerical results can be reproduced using the formulas provided here and in the Supplemental Material; the data are available from the author upon reasonable request.


\bibliography{refs}
\bibliographystyle{utphys}

\clearpage
\onecolumngrid

\setcounter{secnumdepth}{2}
\setcounter{equation}{0}
\setcounter{section}{0}
\setcounter{table}{0}
\setcounter{figure}{0}
\renewcommand{\theequation}{S\arabic{equation}}
\renewcommand{\thetable}{S\arabic{table}}
\renewcommand{\thefigure}{S\arabic{figure}}
\renewcommand{\thesection}{S\arabic{section}}
\renewcommand{\thesubsection}{\Alph{subsection}}
\renewcommand{\theHequation}{S\arabic{equation}}
\renewcommand{\theHtable}{S\arabic{table}}
\renewcommand{\theHfigure}{S\arabic{figure}}

\begin{center}
{\large\bfseries Supplemental Material}
\end{center}
\vspace{0.5em}

This Supplemental Material presents the derivations underlying the main-text results. Section~\ref{sec:master} derives the exact branch evolution and the master formula, Sec.~\ref{sec:theorem1} describes the class of branchwise-definite models and proves Theorem~1, Sec.~\ref{sec:witness} derives the Bargmann witness and the conditional fringe shifts, Sec.~\ref{sec:budget} derives the quantumness budget and analyzes hybrid classical--quantum dynamics, and Sec.~\ref{sec:grav} evaluates the witness for gravity. Equations without the prefix ``S'' refer to the main text. We use $\hbar=c=1$ (except in Sec.~\ref{sec:grav}), the mostly-plus metric, the bracket notation $(T|K|T')$ of Eq.~\eqref{eq:bracket} with $J\equiv(\kappa/2)T$, and the conventions of Ref.~\cite{Matsui:2026skx} for the graviton mode expansion and the radiation amplitudes $\alpha_{T,s}(\bm k)$.

\section{Exact branch evolution and the master formula}
\label{sec:master}

\subsection{Magnus expansion}
For branch $a$ the interaction-picture Hamiltonian is $H_{\rm int}^{(a)}(t)=-A_a(t)$ with $A_a(t)\equiv\int\rmd^3x\,h_{\mu\nu}(t,\bm x)J_a^{\mu\nu}(t,\bm x)$, so that $h[J_a]=\int\rmd t\,A_a(t)$. For a free field the commutator is a $c$-number,
\begin{equation}
[h(x),h(x')]=-i\GPJ(x,x'),\qquad \GPJ(x,x')=i\langle[h(x),h(x')]\rangle,
\label{eq:Sdefs}
\end{equation}
where $\GPJ$ is real and antisymmetric. The retarded and advanced Green functions are $G_R(x,x')=\theta(t-t')\GPJ(x,x')$ and $G_A(x,x')=G_R(x',x)$, so that $\GPJ=G_R-G_A$, and $\GPV\equiv\frac12(G_R+G_A)$ is the time-symmetric part. Smearing with sources gives
\begin{equation}
[h[J],h[J']]=-i(T|\GPJ|T'),\qquad (T|\GPJ|T')=(T|G_R|T')-(T'|G_R|T),
\label{eq:SPeierls}
\end{equation}
which is Eq.~\eqref{eq:comm} and expresses the commutator as the Peierls bracket~\cite{Peierls:1952cb}: by the Kubo formula, $(T|G_R|T')$ is the linear response of $\langle h[J]\rangle$ to the perturbation $-h[J']$, i.e., the retarded field radiated by $T'$ and evaluated on $T$. The Magnus expansion of $U_a=\mathcal{T}\exp[i\int\rmd t\,A_a(t)]$ has
\begin{equation}
\Omega_1=ih[J_a],\qquad
\Omega_2=\frac12\int\rmd t\int^t\rmd t'\,[iA_a(t),iA_a(t')]=\frac{i}{2}\int\rmd^4x\,\rmd^4x'\,\theta(t-t')\,J_a(x)\GPJ(x,x')J_a(x')=\frac{i}{2}(T_a|G_R|T_a),
\end{equation}
and all higher terms vanish because $\Omega_2$ is a $c$-number~\cite{Matsui:2026skx}. Hence $U_a=e^{iS_a}e^{ih[J_a]}$ with $S_a=\frac12(T_a|G_R|T_a)=\frac12(T_a|\GPV|T_a)$, because $(T|\GPJ|T)=0$. Divergent self-energies are absorbed into the matter parameters; only the differences $S_a-S_b$ enter below.

\subsection{Reduced matter state}
With $\rho_{\rm in}=\sum_{ab}\rho^{(0)}_{ab}|a\rangle\langle b|\otimes\rho_h$ and $U=\sum_a|a\rangle\langle a|\otimes U_a$, the reduced matter state is $\sum_{ab}\rho^{(0)}_{ab}M_{ab}|a\rangle\langle b|$ with $M_{ab}=\langle U_b^\dagger U_a\rangle\equiv\Tr(\rho_hU_b^\dagger U_a)$. The Baker--Campbell--Hausdorff formula and Eq.~\eqref{eq:SPeierls} give
\begin{equation}
e^{-ih[J_b]}e^{ih[J_a]}=e^{ih[J_a-J_b]}\,e^{\frac12[h[J_b],h[J_a]]}=e^{ih[J_a-J_b]}\,e^{\frac{i}{2}(T_a|\GPJ|T_b)},
\end{equation}
which yields Eq.~\eqref{eq:master}. The total phase can be rewritten in Feynman--Vernon form~\cite{Feynman:1963fq},
\begin{equation}
S_a-S_b+\tfrac12(T_a|\GPJ|T_b)=(\Delta_{ab}|G_R|\Sigma_{ab}),\qquad \Delta_{ab}=T_a-T_b,\quad \Sigma_{ab}=\tfrac12(T_a+T_b),
\label{eq:SFV}
\end{equation}
i.e., the difference source evaluated on the retarded field of the mean source, so that $M_{ab}=e^{i(\Delta_{ab}|G_R|\Sigma_{ab})}\chi[\Delta_{ab}]$; for the vacuum, this is the influence functional of a Gaussian environment with response kernel $G_R$ and noise kernel $\GH$.

\subsection{Mode representation}
With the transverse-traceless (TT) mode expansion of Ref.~\cite{Matsui:2026skx},
\begin{equation}
ih[J]=\sum_s\int\frac{\rmd^3k}{(2\pi)^3}\Big[\alpha_{T,s}(\bm k)\,a_s^\dagger(\bm k)-\alpha^*_{T,s}(\bm k)\,a_s(\bm k)\Big],
\end{equation}
where $\alpha_{T,s}$ is given by Eq.~(17) of Ref.~\cite{Matsui:2026skx}, so that $e^{ih[J]}=D(\alpha_T)$ is a displacement operator. Then $[ih[J],ih[J']]=\langle\alpha_{T'},\alpha_T\rangle-\langle\alpha_T,\alpha_{T'}\rangle=2i\Img\langle\alpha_{T'},\alpha_T\rangle$, and comparison with Eq.~\eqref{eq:SPeierls} gives Eq.~\eqref{eq:Ealpha}, $(T|\GPJ|T')=2\Img\langle\alpha_{T'},\alpha_T\rangle$. For the vacuum, $\chi[T]=\exp(-\frac12\|\alpha_T\|^2)$, so $\Gamma_{\rm vac}=\frac12(\Delta|\GH|\Delta)=\frac12\|\alpha_{\Delta}\|^2$ for $\Delta=T_a-T_b$, as in Ref.~\cite{Matsui:2026skx}.

\subsection{Gauge invariance and on-shell support}
In de~Donder gauge the retarded Green function of $h_{\mu\nu}$ is $(G_R)_{\mu\nu,\rho\sigma}(x,x')=P_{\mu\nu\rho\sigma}D_R(x-x')$ with $P_{\mu\nu\rho\sigma}=\frac12(\eta_{\mu\rho}\eta_{\nu\sigma}+\eta_{\mu\sigma}\eta_{\nu\rho}-\eta_{\mu\nu}\eta_{\rho\sigma})$ and $D_R$ the massless scalar retarded Green function, and likewise $G_A$, $\GPV$, $\GPJ$, and $\GH$ are $P$ times the scalar functions $D_A$, $D_{\rm PV}$, $D_{\rm PJ}$, and $D_{\rm H}$. Under a gauge transformation $h_{\mu\nu}\to h_{\mu\nu}+\partial_\mu\xi_\nu+\partial_\nu\xi_\mu$, the smeared operator changes by $-\kappa\int\xi_\nu\partial_\mu T^{\mu\nu}=0$, so every quantity in Eq.~\eqref{eq:master} is gauge invariant for conserved sources. The time-symmetric part $D_{\rm PV}=\frac12(D_R+D_A)$ has Fourier transform $\mathcal{P}(1/k^2)$, with no on-shell $\delta(k^2)$ component; contracted with conserved sources it yields the instantaneous Newtonian interaction together with the time-even (conservative) post-Newtonian corrections. The Pauli--Jordan function $D_{\rm PJ}=D_R-D_A$, which determines $\GPJ$, and the vacuum Hadamard function $D_{\rm H}$, which determines $\Gamma_{\rm vac}$, have Fourier transforms proportional to $\mathrm{sgn}(k^0)\delta(k^2)$ and $\delta(k^2)$, respectively. On shell, current conservation $k_\mu\widetilde T^{\mu\nu}=0$ implies the identity~\cite{Weinberg:1972kfs}
\begin{equation}
\widetilde T^*_{\mu\nu}\widetilde T'^{\mu\nu}-\tfrac12\widetilde T^*\widetilde T'=\Lambda_{ij,kl}(\hat{\bm k})\,\widetilde T^{ij*}\widetilde T'^{kl},
\label{eq:Sonshell}
\end{equation}
which is the bilinear form of Eq.~(B1) of Ref.~\cite{Matsui:2026skx}; hence $\GPJ$ and the vacuum noise are given exactly by the two TT helicities, consistent with Eq.~\eqref{eq:Ealpha}. Thus the Newtonian potential enters Eq.~\eqref{eq:master} only through $S_a$, while $\GPJ$ and $\chi$ are radiative.

\section{Branchwise-definite models and Theorem 1}
\label{sec:theorem1}

\subsection{The class $\Cl$}
The class $\Cl$ defined in the main text includes the following models. (i) In semiclassical gravity the field is sourced by $\langle T\rangle$ and branch $a$ acquires $s_a=(T_a|G_R|\langle T\rangle)$ plus self-terms, with no noise. (ii) In Newtonian action at a distance and in quantum-controlled classical fields~\cite{Christodoulou:2018cmk,Martin-Martinez:2022uio}, each branch carries its own classical field sourced by $T_a$, so $s_a$ is a functional of $T_a$ alone; the cross terms between two subsystems reproduce the entangling phase of Refs.~\cite{Bose:2017nin,Marletto:2017kzi}. (iii) A classical stochastic gravitational-wave background is a random free solution $h_{\rm cl}$ independent of the branch. (iv) Di\'osi--Penrose-type models~\cite{Diosi:1989hlx,Penrose:1996cv}, including the Newtonian hybrid of Ref.~\cite{Tilloy:2015zya}, generate a pair potential (contributing to $s_a$) and Gaussian noise coupled linearly to the mass density. Linearity of $h_{\rm cl}$ in $T$ is the defining property of a classical metric perturbation coupled to the stress tensor at the order considered; it is essential, because arbitrary nonlinear branch-dependent phase noise could mimic quantum coherence matrices.

\subsection{A coboundary lemma}
Let $A_{ab}=-A_{ba}$ be real. Then $A_{ab}=f_a-f_b$ for some real $f_a$ if and only if $A_{ab}+A_{bc}+A_{ca}=0$ for all $a,b,c$. Necessity is immediate. For sufficiency, fix a reference branch $1$ and set $f_a\equiv A_{a1}$; then $f_a-f_b=A_{a1}+A_{1b}=A_{ab}-(A_{ab}+A_{b1}+A_{1a})=A_{ab}$. For $A_{ab}=\frac12(T_a|\GPJ|T_b)$, bilinearity and antisymmetry give
\begin{equation}
(T_a|\GPJ|T_b)+(T_b|\GPJ|T_c)+(T_c|\GPJ|T_a)=(T_a-T_c|\GPJ|T_b-T_c),
\end{equation}
so the condition of Theorem~1 is equivalent to $\frac12(T_a|\GPJ|T_b)=f_a-f_b$.

\subsection{Proof of Theorem 1}
Let $W[\phi,\pi]$ be the Wigner functional of $\rho_h$ on the phase space of the radiative sector. Since $h[J]$ is linear in the canonical variables, the Weyl symbol of $e^{ih[J]}$ is $e^{ih_{\rm cl}[T;\phi,\pi]}$, where $h_{\rm cl}[\,\cdot\,;\phi,\pi]$ is the free classical solution with Cauchy data $(\phi,\pi)$. Hence
\begin{equation}
\chi[T]=\int\mathcal{D}\phi\,\mathcal{D}\pi\;W[\phi,\pi]\;e^{ih_{\rm cl}[T;\phi,\pi]}.
\end{equation}
If $W\geq0$ it is a probability measure, and $\chi=\Phi_W$ for the random free field $h_{\rm cl}[\,\cdot\,;\phi,\pi]$ with $(\phi,\pi)\sim W$. Under the commutator condition, the lemma gives $M_{ab}=e^{i(s_a-s_b)}\Phi_W[T_a-T_b]$ with $s_a=S_a+f_a$, which is of the form~\eqref{eq:classical}. $\square$

Remarks. (a) The condition holds automatically for two branches, and for any number of branches of a single interferometer whose difference sources are collinear, $T_a=T_0+\lambda_aX$, since $(\lambda X|\GPJ|\mu X)=0$. (b) Wigner-positive states include all Gaussian states and their mixtures, in particular the phase-averaged relic background used in Ref.~\cite{Matsui:2026skx}; for pure states, $W\geq0$ only for Gaussian states~\cite{HUDSON1974249}. (c) If the noise in $\Cl$ is Gaussian, $\Phi_P[T]=\exp(i\mu[T]-\frac12(T|N|T))$ with $\mu$ linear and $N$ the covariance kernel, every Bargmann invariant of every closed cycle of branches is real and positive, because the arguments $T_{a_k}-T_{a_{k+1}}$ sum to zero and $\mu$ telescopes. For Gaussian graviton states, by contrast, the phase of a cycle equals $\frac12\sum_k(T_{a_k}|\GPJ|T_{a_{k+1}})=\sum_k\Img\langle\alpha_{T_{a_{k+1}}},\alpha_{T_{a_k}}\rangle$, i.e., twice the symplectic area enclosed by the polygon $\{\alpha_{T_{a_k}}\}$, independent of the state.

\section{The Bargmann witness}
\label{sec:witness}

\subsection{Derivation of the witness}
For $T_{ab}=T_{00}+aX+bY$, the differences along the cycle $(00)\to(10)\to(11)\to(01)\to(00)$ are $-X,-Y,X,Y$. The noise factors give $\chi[-X]\chi[-Y]\chi[X]\chi[Y]=|\chi[X]|^2|\chi[Y]|^2$ because $\chi[-T]=\Tr(\rho_h e^{-ih[J]})=\chi[T]^*$. The $S$-phases telescope. With $T\equiv T_{00}$, the brackets $(T_a|\GPJ|T_b)$ in the four commutator phases sum to
\begin{align}
&(T|\GPJ|T+X)+(T+X|\GPJ|T+X+Y)+(T+X+Y|\GPJ|T+Y)+(T+Y|\GPJ|T)\nonumber\\
&\quad=(T|\GPJ|X)+\big[(T|\GPJ|Y)+(X|\GPJ|Y)\big]+\big[(X|\GPJ|T)+(X|\GPJ|Y)\big]+(Y|\GPJ|T)=2(X|\GPJ|Y),
\end{align}
so the cycle carries the phase $(X|\GPJ|Y)$, which is Eq.~\eqref{eq:BQ} together with Eq.~\eqref{eq:Ealpha}. In $\Cl$, the $s$-phases telescope and $\Phi_P[-T]=\Phi_P[T]^*$ for any distribution $P$, giving Eq.~\eqref{eq:BC}. The Bargmann invariant is unchanged by local rephasings $|a\rangle_A\to e^{i\lambda_a}|a\rangle_A$, $|b\rangle_B\to e^{i\mu_b}|b\rangle_B$. For any uncorrelated or pure initial state, the initial four-cycle product is nonnegative; for a correlated mixed initial state, one replaces Eq.~\eqref{eq:Bdef} by the difference between the arguments of the final and initial four-cycle products.

\subsection{Conditional fringe shifts}
Fix $B$ in path $b$. The two branches of $A$ have $\Delta=T_{1b}-T_{0b}=X$ and $\Sigma=T_{00}+\tfrac12X+bY$, so by Eq.~\eqref{eq:SFV} the coherence of $A$ is $M_{1b,0b}=e^{i(X|G_R|T_{00}+X/2+bY)}\chi[X]$ and $\phi_{A|b}=(X|G_R|T_{00}+X/2+bY)+{\rm const}$, where the constant collects the initial phase and $\arg\chi[X]$, neither of which depends on $b$. Hence $\delta\phi_{A|B}=\phi_{A|1}-\phi_{A|0}=(X|G_R|Y)$; likewise, fixing $A$ in path $a$ gives $\Delta=Y$, $\Sigma=T_{00}+aX+\tfrac12Y$, and $\delta\phi_{B|A}=(Y|G_R|X)$, which is Eq.~\eqref{eq:shifts}. Decomposing $G_R=\GPV+\tfrac12\GPJ$,
\begin{equation}
\delta\phi_{A|B}=(X|\GPV|Y)+\tfrac12(X|\GPJ|Y),\qquad
\delta\phi_{B|A}=(X|\GPV|Y)-\tfrac12(X|\GPJ|Y),
\end{equation}
so the conservative part of the response is reciprocal, the commutator part is not, and $\delta\phi_{A|B}-\delta\phi_{B|A}=(X|\GPJ|Y)=\B$. The moduli $|M_{1b,0b}|=|\chi[X]|$ do not depend on $b$. In $\Cl$, $\phi_{A|b}=s_{1b}-s_{0b}+{\rm const}$ and $\phi_{B|a}=s_{a1}-s_{a0}+{\rm const}$, so both conditional shifts equal $s_{11}-s_{10}-s_{01}+s_{00}$.

\subsection{Causal structure and signaling}
(i) If the supports of $X$ and $Y$ are spacelike separated, $(X|\GPJ|Y)=0$ and both retarded responses vanish. (ii) If no point of the support of $X$ lies in the causal future of the support of $Y$ (``$A$ precedes $B$''), then $(X|G_R|Y)=0$, so $\delta\phi_{A|B}=0$ and $\B=-\delta\phi_{B|A}=-(Y|G_R|X)$. (iii) Suppose, in this configuration, that the readout of $A$ is spacelike separated from the beam splitter of $B$, which prepares path probabilities $p_b$. The marginal coherence of $A$ is $\rho^A_{10}\propto\sum_bp_b\,|M_{1b,0b}|\,e^{i\phi_{A|b}}$. In quantum theory $|M_{1b,0b}|$ and $\phi_{A|b}$ are independent of $b$, so $\rho^A$ does not depend on $p_b$. In a model of $\Cl$ that reproduces the observed $\delta\phi_{B|A}\neq0$, reciprocity forces $\delta\phi_{A|B}\neq0$, and the visibility $|\sum_bp_be^{i\phi_{A|b}}|$ depends on the setting $p_b$, enabling signaling. (iv) Under time reversal $\Theta$ about $t_0$, $(\Theta X|\GPJ|\Theta Y)=-(X|\GPJ|Y)$; if both difference sources are invariant, $\B=0$.

\subsection{Nonadditive sources}
Mutual forces deform the trajectories so that $T_{11}-T_{10}-T_{01}+T_{00}\equiv\delta T=O(G)$. The cycle differences become $-X,\,-Y-\delta T,\,X+\delta T,\,Y$, which still sum to zero. For Gaussian noise in $\Cl$, $\mu$ telescopes and $\B_{\Cl}=0$ exactly. For non-Gaussian noise the leading phase is $O(\kappa_3\,\delta T)$, with $\kappa_3$ the third cumulant of $h_{\rm cl}$, i.e., second order in the gravitational coupling and third order in the noise amplitude. In quantum theory the correction to Eq.~\eqref{eq:BQ} is of order $(X|\GPJ|\delta T)$. Both are negligible.

\subsection{Measurement}
Each factor in Eq.~\eqref{eq:Bdef} is a conditional coherence, an expectation value such as $\rho_{00,10}=\langle(|1\rangle\langle0|)_A\otimes|0\rangle\langle0|_B\rangle$, so $\B$ follows from conditional Ramsey fringes of $A$ and $B$ with the partner's path measured, or from cycle tests that estimate Bargmann invariants without tomography~\cite{Oszmaniec:2021rzh}. Since the variance of a phase estimated from $N$ repetitions at visibility $V$ scales as $1/(NV^2)$, resolving $\B$ requires $N\gtrsim(V_XV_Y\B)^{-2}$ up to a factor of order unity.

\section{Quantumness budget and hybrid dynamics}
\label{sec:budget}

\subsection{Uncertainty relation for unitaries}
For unitaries $U,V$ and any state, write $\delta U=U-\langle U\rangle$. Then $\langle[U,V]\rangle=\langle\delta U\delta V\rangle-\langle\delta V\delta U\rangle$, and the Cauchy--Schwarz inequality for the inner product $(A,B)=\Tr(\rho A^\dagger B)$ gives $|\langle\delta U\delta V\rangle|\leq\|\delta U^\dagger\|_\rho\|\delta V\|_\rho=\Delta U\,\Delta V$ with $\Delta U^2=1-|\langle U\rangle|^2$, and similarly for the second term. Hence $\Delta U\,\Delta V\geq\frac12|\langle[U,V]\rangle|$~\cite{Massar:2007akt,Bagchi:2016igu}.

\subsection{Derivation of the budget}
For $U=e^{ih[J_X]}$ and $V=e^{\pm ih[J_Y]}$, Eq.~\eqref{eq:SPeierls} gives $UV=e^{ih[J_X\pm J_Y]}e^{\pm\frac{i}{2}(X|\GPJ|Y)}$ and $VU=e^{ih[J_X\pm J_Y]}e^{\mp\frac{i}{2}(X|\GPJ|Y)}$, so $|\langle[U,V]\rangle|=2|\sin((X|\GPJ|Y)/2)|\,|\chi[X\pm Y]|$, while $\Delta U^2=1-|\chi[X]|^2$ and $\Delta V^2=1-|\chi[Y]|^2$. Using Eq.~\eqref{eq:BQ} gives Eq.~\eqref{eq:budget}. For $\Gamma\ll1$, $1-V^2\simeq2\Gamma$ and $V_{X\pm Y}\simeq1$, giving the first inequality in Eq.~\eqref{eq:budget2}. The second follows from $|\Img\langle\alpha_Y,\alpha_X\rangle|\leq\|\alpha_X\|\|\alpha_Y\|$ and $\Gamma^{\rm vac}_Z=\frac12\|\alpha_Z\|^2$; equality holds if and only if $\alpha_Y=i\lambda\alpha_X$ with real $\lambda$, as for orthogonal-quadrature kicks of a single mode, which therefore saturate Eq.~\eqref{eq:budget2} in the vacuum to leading order.

\subsection{The budget is not a positivity constraint}
Equation~\eqref{eq:budget} relies on the displacement-operator algebra. In particular, the pairs $(00,10)$ and $(11,01)$ share the same modulus $|\chi[X]|$. We drew $2\times10^4$ Gram matrices of four random unit vectors in $\mathbb{C}^6$, which are generic positive coherence matrices with unit diagonal. Labeling the branches $(00),(10),(11),(01)$ as $1,\dots,4$, we find that $38$ of them violate $|\sin(\B/2)|\,|M_{13}|\leq[(1-|M_{12}|^2)(1-|M_{23}|^2)]^{1/2}$. Equation~\eqref{eq:budget} is thus an independent, falsifiable consequence of the quantum-field description.

\subsection{Hybrid classical--quantum dynamics}
Consider a classical field $z=(\phi,\pi)$ coupled to the branch degree of freedom through $H(z)=H_{\rm f}(z)-\sum_a|a\rangle\langle a|\,h_z[T_a]$, evolving with the Aleksandrov--Boucher--Traschen bracket~\cite{Aleksandrov+1981+902+908,Boucher:1988ua}, as in the hybrid dynamics of Refs.~\cite{Diosi:2014ebu,Oppenheim:2018igd}. For the coherence component $\varrho_{ab}(z)$, the bracket $\frac12(\{H,\varrho\}-\{\varrho,H\})_{ab}=\{\frac12(H_{aa}+H_{bb}),\varrho_{ab}\}$ transports $\varrho_{ab}$ along classical trajectories driven by the mean source $\frac12(T_a+T_b)$, while the commutator term multiplies it by the phase $e^{i(h_z[T_a]-h_z[T_b])}$. Along a trajectory, $h_z[T]=h_{\rm free}[T]+(T|G_R|\Sigma_{ab})$, so the accumulated phase is $h_{\rm free}[\Delta_{ab}]+(\Delta_{ab}|G_R|\Sigma_{ab})$, which by Eq.~\eqref{eq:SFV} coincides with the phase of Eq.~\eqref{eq:master}. Such dynamics therefore reproduce $\B=(X|\GPJ|Y)$. Positivity requires additional decoherence and diffusion obeying a trade-off~\cite{Oppenheim:2022xjr,Layton:2023oud}. Gaussian diffusion and decoherence with branch-independent statistics contribute real factors to Bargmann cycles and leave $\B$ unchanged, so these models are distinguished from quantum gravity by the magnitude of decoherence and diffusion, not by $\B$.

\section{Gravitational evaluation}
\label{sec:grav}

\subsection{Quadrupole limit}
For nonrelativistic sources of size and separation $L\ll c\tau$, the difference sources carry no mass monopole (equal energies) and no mass dipole (momentum conservation including recoil), so their leading multipole is the trace-free quadrupole, independent of the origin. Polarizing Eq.~(37) of Ref.~\cite{Matsui:2026skx}, we find that the angular integral of the TT projector between two symmetric trace-free tensors is
\begin{equation}
\int\rmd\Omega_{\bm k}\,\Lambda_{ij,kl}(\hat{\bm k})\,A_{ij}B^*_{kl}=\frac{8\pi}{5}A_{ij}B_{ij}^*.
\end{equation}
Following the steps leading to Eq.~(38) of Ref.~\cite{Matsui:2026skx}, we obtain, with $e^{i\bm k\cdot\bm L}\simeq1$,
\begin{equation}
\langle\alpha_Y,\alpha_X\rangle=\frac{G}{5\pi\hbar c^5}\int_0^\infty\rmd\omega\,\omega^5\,\Delta\widetilde Q^{B*}_{ij}(\omega)\,\Delta\widetilde Q^A_{ij}(\omega),
\end{equation}
which reduces to $\|\alpha\|^2=2\Gamma_{\rm vac}$ for $A=B$. Since $\Delta\widetilde Q(-\omega)=\Delta\widetilde Q(\omega)^*$, $\int_{-\infty}^\infty\rmd\omega\,\omega^5\Delta\widetilde Q^{B*}\Delta\widetilde Q^A=2i\int_0^\infty\rmd\omega\,\omega^5\Img(\Delta\widetilde Q^{B*}\Delta\widetilde Q^A)$, and with $\widetilde f(\omega)=\int\rmd t\,e^{i\omega t}f(t)$ and Parseval's identity the left-hand side equals $2\pi i\int\rmd t\,\Delta Q^B_{ij}\,\rmd^5\Delta Q^A_{ij}/\rmd t^5$. Hence
\begin{equation}
\B=2\Img\langle\alpha_Y,\alpha_X\rangle=\frac{2G}{5\pi\hbar c^5}\int_0^\infty\rmd\omega\,\omega^5\Img\big(\Delta\widetilde Q^{B*}_{ij}\Delta\widetilde Q^A_{ij}\big)=\frac{2G}{5\hbar c^5}\int\rmd t\,\Delta Q^B_{ij}\,\frac{\rmd^5\Delta Q^A_{ij}}{\rmd t^5},
\end{equation}
which is Eq.~\eqref{eq:quad}. Integration by parts shows antisymmetry under $A\leftrightarrow B$. The structure $\Delta Q^B_{ij}\,\rmd^5\Delta Q^A_{ij}/\rmd t^5$ is the interaction of $B$ with the Burke--Thorne radiation-reaction potential $\Phi_{\rm RR}=(G/5c^5)x^ix^j\,\rmd^5Q_{ij}/\rmd t^5$ generated by $A$~\cite{Burke:1970wx,Thorne:1969rba,Blanchet:2013haa}, the first time-odd term of the near-zone metric.

\subsection{Gaussian trajectory model}
In the model of Ref.~\cite{Matsui:2026skx}, $\Delta Q^A_{ij}(t)=m_Ad_A^2e^{-t^2/\tau^2}S^A_{ij}$ and $\Delta Q^B_{ij}(t)=m_Bd_B^2e^{-(t-\Delta t)^2/\tau^2}S^B_{ij}$ with $S_{ij}=e_ie_j-\frac13\delta_{ij}$, so that $S^A_{ij}S^B_{ij}=(\bm e_A\cdot\bm e_B)^2-\frac13$. Their transforms are $\Delta\widetilde Q^A=\sqrt\pi\,m_Ad_A^2\tau\,e^{-\omega^2\tau^2/4}S^A$ and $\Delta\widetilde Q^B=\sqrt\pi\,m_Bd_B^2\tau\,e^{-\omega^2\tau^2/4}e^{i\omega\Delta t}S^B$, and
\begin{equation}
\B=-\frac{2Gm_Am_Bd_A^2d_B^2}{5\hbar c^5\tau^4}\Big[(\bm e_A\cdot\bm e_B)^2-\tfrac13\Big]I(\delta),\qquad
I(\delta)\equiv\int_0^\infty\rmd\nu\,\nu^5e^{-\nu^2/2}\sin(\nu\delta).
\end{equation}
From $\int_0^\infty\rmd\nu\,e^{-\nu^2/2}\cos(\nu\delta)=\sqrt{\pi/2}\,e^{-\delta^2/2}$ and $\frac{\rmd}{\rmd\delta}[\mathrm{He}_n(\delta)e^{-\delta^2/2}]=-\mathrm{He}_{n+1}(\delta)e^{-\delta^2/2}$, one finds $I(\delta)=\sqrt{\pi/2}\,\mathrm{He}_5(\delta)e^{-\delta^2/2}=\sqrt{\pi/2}\,(\delta^5-10\delta^3+15\delta)e^{-\delta^2/2}$, which is Eq.~\eqref{eq:gauss}. Its global maximum is $|I|_{\max}=7.248$ at $\delta=0.617$, and $|I(\delta)|\leq\int_0^\infty\rmd\nu\,\nu^5e^{-\nu^2/2}=8$ for all $\delta$. With $\Gamma_{\rm vac}=\frac{8}{15}Gm^2d^4/\hbar c^5\tau^4$ the budget ratio is
\begin{equation}
\frac{|\B|}{4\sqrt{\Gamma^{\rm vac}_X\Gamma^{\rm vac}_Y}}=\frac{3}{16}\Big|(\bm e_A\cdot\bm e_B)^2-\tfrac13\Big|\,|I(\delta)|\leq\frac{3}{16}\cdot\frac23\cdot8=1,
\end{equation}
consistent with Eq.~\eqref{eq:budget2}; the maximum attained is $0.906$. For identical interferometers with parallel axes, $|\B|_{\max}=\frac{4}{15}|I|_{\max}\,Gm^2d^4/\hbar c^5\tau^4=3.62\,\Gamma_{\rm vac}$. With the parameters and $\Gamma_{\rm vac}$ of Table~I of Ref.~\cite{Matsui:2026skx}, this gives $|\B|_{\max}\simeq3.2\times10^{-60}$ for the BMV/QGEM parameters and $1.4\times10^{-88}$ (MAQRO) to $9.0\times10^{-70}$ ($^{87}$Rb interferometer) for the other platforms, i.e., $N_{\min}\simeq|\B|_{\max}^{-2}\sim10^{119}$--$10^{176}$ repetitions.

\end{document}